\documentclass[a4paper,12pt]{article}
\title{
\begin{flushright}
{\normalsize OCU-PHYS-205}
\end{flushright}
  Perturbative out of equilibrium quantum field theory beyond the gradient 
  approximation and generalized Boltzmann equation.
  }
\author{
  H. Ozaki\footnote{E-mail address: hozaki@sci.osaka-cu.ac.jp}\\
  \textit{Graduate school of science, Osaka City University,}\\
  \textit{Sumiyoshi-ku, Osaka 558-8585, Japan}
  }
\date{(November, 2003)}

\begin{document}
\maketitle
\begin{abstract}
Using the closed-time-path formalism, we construct perturbative frameworks, in terms of quasiparticle picture, for studying quasiuniform relativistic quantum field systems near equilibrium and non-equilibrium quasistationary systems.
We employ the derivative expansion and take in up to the second-order term, i.e., one-order higher than the gradient approximation.
After constructing self-energy resummed propagator, we formulated two kind of mutually equivalent perturbative frameworks: 
The first one is formulated on the basis of the ``bare'' number density function, and the second one is formulated on the basis of ``physical'' number density function.
In the course of construction of the second framework, the generalized Boltzmann equations \textit{directly} come out, which describe the evolution of the system.
\end{abstract}

\section{Introduction}\label{sec:Intro}
Ultrarelativistic heavy-ion collision experiments have begun at the BNL Relativistic Heavy Ion Collider (RHIC) and will soon start at the CERN Large-Hadron Collider (LHC) in expectation of producing quark-gluon plasma (QGP)~\cite{Bjorken,LeBellac}. \ 
A produced system in such a experiment is nonequilibrium and evolves into confined (or hadronic) and chiral-symmetry breaking phase through phase transitions.
For dealing with nonequilibrium quantum-field systems, several formalisms are available.
Among those are the closed-time-path (CTP) formalism~\cite{Schwinger,Chou} and the nonequilibrium thermo field dynamics~\cite{Umezawa}. \ 
In this paper, we employ the former.

Throughout this paper, we are interested in quasiuniform relativistic quantum field systems near equilibrium and nonequilibrium quasistationary systems, which we simply refer to as ``out of equilibrium'' systems.
The out of equilibrium systems are characterized by two different spacetime scales~\cite{Chou}. \ 
The first one is microscopic or quantum-field theoretical scale, which characterizes the range of radiative corrections to the reactions taking place in the system.
The second scale is macroscopic or statistical scale, which measures the relaxation of the system.
For a weak coupling theory, in which we are interested, the former scale is much smaller than the latter one.

The ``two-scale structure'' mentioned above is implemented as follows.
Let $G(x,y)$ be a generic two-point function.
$G(x,y)$, with $x-y$ fixed, does not change appreciably in $(x+y)/2$.
The relative coordinates $x-y$ are chosen as the microscopic coordinates while $X \equiv (x+y)/2$ are chosen as the macroscopic coordinates.
A Wigner transformation, i.e., a Fourier transform with respect to $x-y$, yields
\begin{equation}
G(x,y) \equiv \int \frac{d^4 P}{(2\pi)^4} e^{-iP\cdot(x-y)} G(X;P)
~~~~~ \left[ X \equiv (x+y)/2 \right]
\label{eq:FourierXP}
\end{equation}
$\left[ P^\mu = (p^0,\mathbf{p}) \right]$.
Then $P^\mu$ in Eq.(\ref{eq:FourierXP}) can be regarded as the momentum of quasiparticle participating in the microscopic reaction under consideration.
Because of the weak $X$-dependence of $G(X;P)$, the derivative expansion makes sense,
\begin{equation}
G(X;P) = 
\left[
1 + (X-Y)^\mu \partial_{Y^\mu} 
+ \frac{1}{2} (X-Y)^\mu(X-Y)^\nu \partial_{Y^\mu} \partial_{Y^\nu} 
+ \dots
\right]
G(Y;P),
\label{eq:Bibuntenkai}
\end{equation}
where $\partial_{Y^\mu} = \partial/\partial Y^\mu$.
We refer to the first term, $G(Y;P)$, on the right-hand side (RHS) as the leading or zeroth-order part (term), to the second term to the gradient or first-order part (term), and to the third term as the second-order part (term).
Most works reported as far ( see, e.g.,~\cite{Blaizot,Niegawa,NieBoseFermi}) on out-of-equilibrium perturbation theory are carried out within the gradient approximation, i.e., the approximation of taking up to the gradient term in Eq.(\ref{eq:Bibuntenkai}).

The purpose of this paper is to frame perturbative frameworks beyond the gradient approximation by taking up to the second-order term on the RHS of Eq.(\ref{eq:Bibuntenkai}).
We construct two mutually equivalent perturbative frameworks.
One framework is constructed in terms of the ``bare''-number density function, and the other, which is called physical-$N$ scheme, is constructed in terms of the ``renormalized''-number density function.
For definiteness, we take up an out-of-equilibrium complex-scalar quantum field system\footnote{This system has been dealt with in~\cite{Niegawa} to the gradient approximation. Fermion and gauge-boson systems have been treated in~\cite{NieBoseFermi}.}.
Generalization to other theories is important.
In this relation, it has been pointed out that a care should be taken~\cite{Blaizot} when one applies the derivative expansion to gauge theories.

The perturbative framework to be constructed accompanies the generalized Boltzmann equation (GBE) for the number density of quasiparticles.
The framework allows us to compute any reaction rate by using the reaction rate formula~\cite{oj1999}.
Substituting the computed net production rates of quasiparticles into the GBE, one can determine the number densities as functions of macroscopic spacetime coordinates $X^\mu$, which describes the evolution of the system.

The plan of the paper is as follows.
In Sec.~\ref{sec:CTP}, on the basis of the CTP formalism supplemented with a quasiparticle representation, we construct the free quasiparticle propagator up to the second-order term in the derivative expansion, Eq.(\ref{eq:Bibuntenkai}).
In Sec.~\ref{sec:Resum}, we construct the self-energy resummed propagator and observe that large contributions come out in general.
In Sec.~\ref{sec:bareNscheme}, after introducing ``physical'' number density, we briefly describe ``the bare-$N$ scheme.''
In Sec.~\ref{sec:Boltzmann}, imposing the condition that no large contributions appears, we construct a ``healthy'' perturbative framework (the physical-$N$ scheme).
It is shown that, on the energy shell, the condition turns out to the generalized Boltzmann equation.
Furthermore we impose the condition that, on the energy shell (with $\mathbf{p}=0$), the medium effect as well as the quantum correction to the ``mass'' vanish, which leads to a gap equation.
In Sec.~\ref{sec:Discussion}, the similarity between the bare-$N$ scheme and the physical-$N$ scheme is discussed.
Sec.~\ref{sec:Summary} is devoted to summary.

\section{Closed-time-path Formalism}\label{sec:CTP}
\subsection{Preliminaries}\label{ssec:Prelim}
The Lagrangian (density) of the complex-scalar field theory reads
\begin{equation}
\mathcal{L} = 
-\phi^\dagger(x) \left( \partial_x^2 + m^2 \right) \phi(x)
-\frac{\lambda}{4}\left( \phi^\dagger(x) \phi(x) \right)^2.
\label{eq:Lag}
\end{equation}
To obtain an efficient perturbative scheme, we introduce a weakly $x$-dependent mass function $M(x,y)$~\cite{Hatsuda} and write $\mathcal{L}$ as 
\begin{eqnarray}
\mathcal{L} &=& \mathcal{L}_0 + \mathcal{L}_{c_1} + \mathcal{L}_{\mathrm{int}},
\label{eq:LagAll}\\
\mathcal{L}_0 &=& 
-\phi^\dagger(x) 
\int d^4y
\left[
\delta^4(x-y) \partial^2_y + \mathcal{M}^2(x,y)
\right]
\phi(y),
\label{eq:Lag0}\\
\mathcal{L}_{c_1} &=& 
\phi^\dagger(x) 
\int d^4y
\left[
\mathcal{M}^2(x,y) - m^2 \delta(x-y)
\right]
\phi(y),
\label{eq:LagC1}\\
\mathcal{L}_{\mathrm{int}} &=& 
-\frac{\lambda}{4}\left( \phi^\dagger \phi \right)^2,
\label{eq:LagInt}
\end{eqnarray}
where 
\begin{eqnarray}
\mathcal{M}^2(x,y) &=& 
\theta \left( i(\partial_{x^0}- \partial_{y^0}) \right) 
M^2_+ \left( X \right)
+ \theta \left( -i(\partial_{x^0} - \partial_{y^0}) \right)
M^2_- \left( X \right)
\label{eq:M^2(x,y)}\\
&=&
\int \frac{d^4P}{(2\pi)^4} e^{-iP \cdot (x-y)} \mathcal{M}^2 (X;p_0)
 ~~~~~ \left( X = (x+y)/2 \right),
\label{eq:M^2(x,y)Wig}\\
\mathcal{M}^2(X;p_0) &=& \theta(p_0) M^2_+(X) + \theta(-p_0) M^2_-(X).
\label{eq:M^2Xp0}
\end{eqnarray}
Here $M_+ (M_-)$ is the effective mass of a quasiparticle (an anti-quasiparticle).
How to determine $M^2_\pm$ is discussed in Sec.~\ref{ssec:Gap}.
Note that $(x+y)/2$ in $M^2_\pm((x+y)/2)$ is a macroscopic spacetime coordinates.
It is well known that an introduction of ultraviolet renormalization counter terms in vacuum theory is sufficient for the out of equilibrium theory to be renormalized.
So we do not explicitly write down those counter terms.
We assume, for simplicity, that the density matrix $\rho$ commutes with the charge operator Q: $\left[ \rho, Q \right] = 0$ (Generalization to the case $\left[ \rho, Q \right] \ne 0$ is straightforward.).

The CTP formalism is formulated~\cite{Chou} by introducing an oriented closed-time path $C (= C_1 + C_2)$ in a complex time plane.
$C_1$ runs along real $x_0$-axis from $-\infty$ to $+\infty-i0^+$, and $C_2$ comes back from $+\infty-i0^+$ to $-\infty-2i0^+$.
The real-time formalism is achieved by doubling every degree of freedom
$\phi \rightarrow (\phi_1, \phi_2)$, where $\phi_1 = \phi(x_0,\mathbf{x})$ with $x_0 \in C_1$ is called a type-1 field and $\phi_2 = \phi(x_0,\mathbf{x})$ with $x_0 \in C_2$ is called a type-2 field.
The classical contour action is written as
\begin{eqnarray}
\int_{C} dx_0 \int d\mathbf{x} \mathcal{L}(\phi(x), \phi^\dagger(x)) &=& 
\int_{-\infty}^{+\infty} dx_0 \int d\mathbf{x} \hat{\mathcal{L}}(x),
\label{eq:ClassicAct}\\
\hat{\mathcal{L}} \equiv 
\mathcal{L}(\phi_1,\phi_1^\dagger)
&-&\mathcal{L}(\phi_2,\phi_2^\dagger).
\label{eq:hatLag}
\end{eqnarray}
$\hat{\mathcal{L}}$ is called a hat-Lagrangian~\cite{Umezawa}. \ 
Introducing a doublet notation,
\begin{equation}
\hat{\phi}^\dagger = (\phi_1^\dagger, \phi_2^\dagger), ~~~~~
\hat{\phi} = 
\left( 
\begin{array}{c}
\phi_1\\
\phi_2
\end{array} 
\right), ~~~~~
\hat{\tau}_3 = 
\left( 
\begin{array}{lr}
1 & 0 \\
0 & -1 
\end{array} 
\right),
\label{eq:hathathat}
\end{equation}
$\hat{\mathcal{L}}_0$ and $\hat{\mathcal{L}}_{c1}$ may be written in the forms,
\begin{eqnarray}
\hat{\mathcal{L}}_0 &=& 
- \hat{\phi}^\dagger (x)
\int d^4y
\left[
\delta^4(x-y) \partial^2_y + \mathcal{M}^2(x,y)
\right]
\hat{\tau}_3 \hat{\phi}(y),
\label{eq:hatLag_0}\\
\hat{\mathcal{L}}_{c_1} &=& 
\hat{\phi}^\dagger(x)
\int d^4y
\left[
\mathcal{M}^2(x,y) - m^2 \delta^4(x-y)
\right]
\hat{\tau}_3 \hat{\phi}(y).
\label{eq:hatLag_c1}
\end{eqnarray}

The vertex factors may be read off from Eqs.(\ref{eq:LagInt}), (\ref{eq:hatLag}) and (\ref{eq:hatLag_c1}).
The vertex factor for a four-point vertex is extracted from $\hat{\mathcal{L}}_{\mathrm{int}}$:
\begin{equation}
i\hat{V} = 
\left( 
\begin{array}{lr}
-i\lambda & 0\\
0 & i\lambda
\end{array} 
\right)
= -i\lambda \hat{\tau}_3.
\label{eq:Vertex}
\end{equation}
Here, $iV_{11} = -i\lambda \left[ iV_{22}=i\lambda \right]$ is the vertex factor for a vertex of type-1 [type-2] fields.
For a two-point vertex coming from $\hat{\mathcal{L}}_{c1}$, we obtain 
\begin{equation}
i\hat{V}_{c1}(x,y) 
= 
i \left( \mathcal{M}^2(x,y) - m^2 \delta^4(x-y) \right) \hat{\tau}_3.
\label{eq:Vertex_c1}
\end{equation}
In momentum space, $i\hat{V}_{c1}$ becomes
\begin{equation}
i \hat{V}_{c1}(X;P)
=
i
\left[
\mathcal{M}^2(X;p_0) - m^2
\right]
\hat{\tau}_3,
\label{eq:hatVc1XP}
\end{equation}
Throughout this paper we do not deal with initial-correlations (see, e.g.,~\cite{Chou}).

\subsection{Free quasiparticle propagator}\label{ssec:BarePrp}
In this subsection, we derive the form for the free quasiparticle propagator with the help of quasiparticle representation~\cite{Chou,Umezawa,Niegawa}. \

\subsubsection{Preliminary}\label{sssec:Prelim}
Following standard procedure~\cite{Chou}, the four kinds of propagators emerge:
\begin{eqnarray}
\Delta_{11}(x,y) &\equiv& -i \mathrm{Tr}
\left[ T \left\{ \phi_1(x) \phi^\dagger_1(y) \right\} \rho \right],
\label{Delta_11(x,y)}\\
\Delta_{12}(x,y) &\equiv& -i \mathrm{Tr}
\left[ \phi^\dagger_2(y) \phi_1(x) \rho \right],
\label{Delta_12(x,y)}\\
\Delta_{21}(x,y) &\equiv& -i \mathrm{Tr}
\left[ \phi_2(x) \phi^\dagger_1(y) \rho \right],
\label{Delta_21(x,y)}\\
\Delta_{22}(x,y) &\equiv& -i \mathrm{Tr}
\left[ \bar{T} \left\{ \phi_2(x) \phi^\dagger_2(y) \right\} \rho \right],
\label{Delta_22(x,y)}
\end{eqnarray}
where $\rho$ is the density matrix, and $T (\bar{T})$ is the time-ordering (antitime-ordering) symbol, $\phi_1, \phi_1^\dagger, \phi_2$ and $\phi_2^\dagger$ are the interaction-picture fields.
At the end of calculation we set $\phi_1 = \phi_2$~\cite{Chou}. \ 
Let us introduce a matrix propagator $\hat{\Delta}(x,y)$, where the ``caret'' denotes the $2 \times 2$ matrix, whose $(ij)$ component is $\Delta_{ij}(x,y)$.
The matrix self-energy $\hat{{\Sigma}}(x,y)$ is defined similarly.
$\hat{\Delta}(x,y)$ is defined by (c.f. Eq.(\ref{eq:hatLag_0}))
\begin{eqnarray}
\int d^4z 
\left( \delta^4(x-z) \partial_z^2 + \mathcal{M}^2(x,z) \right) 
\hat{\Delta}(z,y) 
&=& 
\int d^4z
\hat{\Delta}(x,z) \left( \overleftarrow{\partial^2_z} \delta^4(z-y) + 
\mathcal{M}^2(z,y) 
\right) 
\nonumber\\
&=& 
- \hat{\tau}_3 \delta^4(x-y).
\label{eq:yama}
\end{eqnarray}
In the sequel, we use the short-hand notation $H = F \cdot G$, whose ``$(x,y)$ component'' is
\begin{equation}
H(x,y) = \left[ F \cdot G \right](x,y) = \int d^4z F(x,z)G(z,y).
\label{eq:dotSeki}
\end{equation}

\subsubsection{Quasiparticle representation}\label{sssec:Quasirepr}
Motivated by equilibrium thermal field theory (ETFT), we introduce, as usual~\cite{Umezawa}, a quasiparticle representation for $\hat{\Delta}(x,y)$:
\begin{eqnarray}
\hat{\Delta}(x,y) &=& \hat{B}_L \cdot \hat{\Delta}_{RA} \cdot \hat{B}_R,
\label{eq:hatDelta(x,y)}\\
\hat{\Delta}_{RA}(x,y) &=& \mathrm{diag}\left( \Delta_R, -\Delta_A \right),
\label{eq:hatDelta_RA}\\
\hat{B}_L(x,y) &=& 
\left(
\begin{array}{cc}
\delta(x-y) & f(x,y)\\
\delta(x-y) & \delta(x-y) + f(x,y)
\end{array} 
\right),
\label{eq:hatB_L(x,y)}\\
\hat{B}_R(x,y) &=& 
\left(
\begin{array}{cc}
\delta(x-y) + f(x,y) & f(x,y)\\
\delta(x-y)          & \delta(x-y)
\end{array} 
\right).
\label{eq:hatB_R(x,y)}
\end{eqnarray}
Here $\Delta_R (\Delta_A)$ is the retarded (advanced) propagator.
As will be seen in Sec.\ref{ssec:NumberDensity}, $f(x,y)$ is related to the number density.
In particular, in equilibrium thermal field theory, $f(x,y) = f(x_0 - y_0) \delta( \mathbf{x} - \mathbf{y} )$ and 
\begin{equation}
\int d^4x e^{iP \cdot x} f(x) = 
\epsilon(p_0) \tilde{n}_{\epsilon(p_0)}(|p_0|) - \theta(-p_0)
\label{eq:ETFTf}
\end{equation}
with $\tilde{n}_+ (\tilde{n}_-)$ the number density of a quasiparticle (an anti-quasiparticle).

\subsubsection{Retarded (advanced) propagator $\Delta_{R(A)}$}\label{sssec:RAprop}
$\Delta_R(x,y) \left[ \Delta_A(x,y) \right]$ is obtained by solving
\begin{eqnarray}
&&\int d^4z
\left( \delta^4(x-z) \partial_z^2 + \mathcal{M}^2(x,z) \right) 
\Delta_{R(A)}(z,y) 
\nonumber\\
& &= 
\int d^4z
\Delta_{R(A)}(x,z) \left( \overleftarrow{\partial^2_z} \delta^4(z-y)  
+ \mathcal{M}^2(z,y) \right)
\nonumber\\
& &=
-\delta^4(x-y),
\label{eq:KernelDelta_RA(x,y)}
\end{eqnarray}
under the retarded (advanced) boundary condition.
Wigner-transforming and taking up to the second-order terms in the derivative expansion, we obtain, with obvious notation,
\begin{eqnarray}
\Delta_{R(A)}(X;P) &\simeq& 
\Delta_{R(A)}^{(0)}(X;P) + \Delta_{R(A)}^{(2)}(X;P),
\label{eq:Delta_R/A(X;P)}\\
\Delta_{R}^{(0)}(X;P) &=& 
\left( \Delta_{A}^{(0)}(X;P) \right)^\ast
= \frac{1}{P^2 - \mathcal{M}^2(X;p_0) + i \epsilon (p_0) 0^+},
\label{eq:Delta_R/A^(0)}\\
\Delta_{R(A)}^{(2)}(X;P) &=&
\frac{1}{2} \bigg[ 
\partial_X^2 \mathcal{M}^2(X;p_0) \left( \Delta_{R(A)}^{(0)} \right)^3
- 2 P^\mu P^\nu \partial_{X^\mu} \partial_{X^\nu} \mathcal{M}^2(X;p_0) 
\left( \Delta_{R(A)}^{(0)} \right)^4
\nonumber\\
& &~~~~
+ \partial_{X^\mu} \mathcal{M}^2(X;p_0) \partial_{X_\mu} \mathcal{M}^2(X;p_0)
\left( \Delta_{R(A)}^{(0)} \right)^4
 \bigg],
\label{eq:Delta_R/A^(2)}
\end{eqnarray}
where $\mathcal{M}^2(X;p_0)$ is as in Eq.(\ref{eq:M^2Xp0}).
It should be noted that no gradient term appears.

\subsubsection{Free quasiparticle propagator $\hat{\Delta}$}\label{sssec:Bareprop}
Wigner transformation of Eq.(\ref{eq:hatDelta(x,y)}) yields
\begin{eqnarray}
\hat{\Delta}(X;P) &\simeq& 
\left( 
\begin{array}{cc}
\Delta_R & 0
\\
\Delta_R - \Delta_A & -\Delta_A
\end{array}
 \right)
+ \left( \Delta_R - \Delta_A \right) f \hat{A}_+
\nonumber\\
& &+ \frac{1}{2i} \left\{ \Delta_R^{(0)} + \Delta_A^{(0)}, f \right\} \hat{A}_+
- \frac{1}{8} 
\left\{ \left\{ \Delta_R^{(0)} - \Delta_A^{(0)}, f \right\} \right\} 
\hat{A}_+,
\label{eq:hatDelta(X;P)}\\
\hat{A}_\pm &=& 
\left( 
\begin{array}{cc}
1  &  \pm 1  \\
\pm 1  &  1  
\end{array}
\right),
\label{eq:A_+-}\nonumber
\end{eqnarray}
where $\Delta_{R(A)} = \Delta_{R(A)}(X;P)$, and $f = f(X;P)$.
$\left\{ \cdots, \cdots \right\}$ and $\left\{ \left\{ \cdots, \cdots \right\} \right\}$ are defined as
\begin{eqnarray}
\left\{ A(X;P), B(X;P) \right\} &\equiv&
\frac{\partial A}{\partial X_\mu} \frac{\partial B}{\partial P^\mu}
- \frac{\partial A}{\partial P^\mu} \frac{\partial B}{\partial X_\mu},
\label{eq:Poisson}\\
\left\{ \left\{ A(X;P), B(X;P) \right\} \right\} &\equiv&
-2\frac{\partial^2 A}{\partial X_\mu \partial P^\nu}
\frac{\partial^2 B}{\partial P^\mu \partial X_\nu}
\nonumber\\
& &+\frac{\partial^2 A}{\partial X_\mu \partial X_\nu}
\frac{\partial^2 B}{\partial P^\mu \partial P^\nu}
+\frac{\partial^2 A}{\partial P^\mu \partial P^\nu}
\frac{\partial^2 B}{\partial X_\mu \partial X_\nu},
\label{eq:Poisson2}
\end{eqnarray}
respectively.
Straightforward computation using Eq.(\ref{eq:Delta_R/A^(0)}) yields
\begin{eqnarray}
\left\{ \Delta_R^{(0)} + \Delta_A^{(0)}, f \right\} &=&
\left\{ f, P^2 - \mathcal{M}^2(X;p_0) \right\} 
\left( {\Delta_R^{(0)}}^2 + {\Delta_A^{(0)}}^2 \right),
\label{eq:[D_R+D_A,f]}\\
\left\{ \left\{ \Delta_R^{(0)} - \Delta_A^{(0)}, f \right\} \right\} &=& 
\left[
\partial_{X^\mu} \partial_{X^\nu} \mathcal{M}^2(X;p_0)
\partial_{P_\mu} \partial_{P_\nu} f
-2 \partial_X^2 f
\right] \left( {\Delta_R^{(0)}}^2 - {\Delta_A^{(0)}}^2 \right)
\nonumber\\
& &+ \bigg[ 
8 P^\mu \partial_{X^\nu} \mathcal{M}^2(X;p_0) 
\partial_{P_\nu} \partial_{X^\mu}f
\nonumber\\
& &~~~~
+ 2 \left( \partial_{X^\mu} \mathcal{M}^2(X;p_0) \right) 
\left( \partial_{X^\nu} \mathcal{M}^2(X;p_0) \right)
\partial_{P_\mu}\partial_{P_\nu} f
\nonumber\\
& &~~~~
+ 8 P^\mu P^\nu \partial_{X^\mu} \partial_{X^\nu} f
 \bigg] \left( {\Delta_R^{(0)}}^3 - {\Delta_A^{(0)}}^3 \right).
\label{eq:[[D_R-D_A,f]]}
\end{eqnarray}
We may rearrange Eq.(\ref{eq:hatDelta(X;P)}) in the form
$\hat{\Delta}(X;P) \simeq \hat{\Delta}^{(0)} + \hat{\Delta}^{(1)}
+ \hat{\Delta}^{(2)}$, where $\hat{\Delta}^{(j)} (j=0,1,2)$ stands for the $j$th-order term in the derivative expansion:
\begin{eqnarray}
\hat{\Delta}^{(0)}(X;P) &=& 
\left( 
\begin{array}{cc}
\Delta_R^{(0)} & 0
\\
\Delta_R^{(0)} - \Delta_A^{(0)} & -\Delta_A^{(0)}
\end{array}
 \right)
+ \left( \Delta_R^{(0)} - \Delta_A^{(0)} \right) f \hat{A}_+ ,
\label{eq:hatDelta0}\\
\hat{\Delta}^{(1)}(X;P) &=& 
\frac{1}{2i}
\left\{ \Delta_R^{(0)} + \Delta_A^{(0)}, f \right\} \hat{A}_+,
\label{eq:hatDelta1}\\
\hat{\Delta}^{(2)}(X;P) &=& 
\left( 
\begin{array}{cc}
\Delta_R^{(2)} & 0
\\
\Delta_R^{(2)} - \Delta_A^{(2)} & -\Delta_A^{(2)}
\end{array}
 \right)
+ \left( \Delta_R^{(2)} - \Delta_A^{(2)} \right) f \hat{A}_+ 
\nonumber\\
& &
-\frac{1}{8} 
\left\{ \left\{ \Delta_R^{(0)} - \Delta_A^{(0)}, f \right\} \right\} 
\hat{A}_+.
\label{eq:hatDelta2}
\end{eqnarray}
Eq.(\ref{eq:yama}) with Eqs.(\ref{eq:hatDelta0})-(\ref{eq:hatDelta2}) yields
\begin{eqnarray}
\frac{1}{2i} 
\left[ 
2P^\mu \partial_{X^\mu}f(X;P) 
+ \partial_{P_\mu}f(X;P) \partial_{X^\mu} \mathcal{M}^2(X;p_0) \right]
\left(  
{\Delta_R^{(0)}}^2 + {\Delta_A^{(0)}}^2
\right)
\nonumber\\
- \frac{1}{8}
\left\{ \left\{  
\Delta_R^{(0)} - \Delta_A^{(0)} , f(X;P)
\right\} \right\}
= 0.
\label{eq:freeBoltz}
\end{eqnarray}
Except the fact that a medium effect is taken in $\mathcal{M}^2(X;p_0)$, Eq.(\ref{eq:freeBoltz}) leads to a ``free Boltzmann equation.'' (see remark about after Eq.(\ref{eq:hatB_R(x,y)}) and Sec.~\ref{ssec:Boltzmann} below.)

\section{Resummation of the self-energy}\label{sec:Resum}
In this section, we construct and analyze the self-energy resummed propagator, and disclose the existence of a large contribution, which spoils the consistency of perturbative framework.

\subsection{Self-energy resummed propagator}\label{ssec:ResumPropagator}
Let $\hat{\Sigma} = \hat{\Sigma}(x,y)$ be a self-energy matrix.
The $\hat{\Sigma}$-resummed propagator $\hat{G}$ obeys the Dyson equation,
\begin{equation}
\hat{G}(x,y) = \hat{\Delta} + \hat{\Delta} \cdot \hat{\Sigma} \cdot \hat{G}
=
\hat{G} \cdot \hat{\Sigma} \cdot \hat{\Delta} + \hat{\Delta}.
\label{eq:Schwinger-Dyson}
\end{equation}
Multiplying $\hat{B}_L^{-1} (\hat{B}_R^{-1})$ from the left (right) of both sides of Eq.(\ref{eq:Schwinger-Dyson}), we have 
\begin{equation}
\underline{\hat{G}}(x,y) = 
\hat{\Delta}_{RA} + 
\hat{\Delta}_{RA} \cdot \underline{\hat{\Sigma}} \cdot \underline{\hat{G}}
= 
\hat{\Delta}_{RA} + 
\underline{\hat{G}} \cdot \underline{\hat{\Sigma}} \cdot \hat{\Delta}_{RA},
\label{eq:ulhatG}
\end{equation}
where use has been made of Eq.(\ref{eq:hatDelta(x,y)}) and 
\begin{eqnarray}
\underline{\hat{G}}(x,y) &=& \hat{B}_L^{-1} \cdot \hat{G} \cdot \hat{G}_R^{-1}
\label{eq:haha}\\
\underline{\hat{\Sigma}}(x,y) &=& 
\hat{B}_R \cdot \hat{\Sigma} \cdot \hat{B}_L 
= 
\left(  
\begin{array}{cc}
\Sigma_R  &  \Sigma_K  \\
0  &  -\Sigma_A
\end{array}
\right).
\label{eq:ulhatSigma(x,y)}
\end{eqnarray}
Here
\begin{eqnarray}
\Sigma_R(x,y) &=& \Sigma_{11} + \Sigma_{12},~~
\Sigma_A(x,y) = - \left( \Sigma_{22} + \Sigma_{12} \right),
\label{eq:Sigma_R/A(x,y)}\\
\Sigma_K(x,y) &=& 
\Sigma_{12} + \Sigma_{12} \cdot f - \Sigma_{21} \cdot f 
+ \Sigma_A \cdot f - f \cdot \Sigma_A.
\label{eq:Sigma_K(x,y)}
\end{eqnarray}
In obtaining Eqs.(\ref{eq:ulhatSigma(x,y)})-(\ref{eq:Sigma_K(x,y)}), we have used the relation $\sum_{i,j=1}^{2} \Sigma_{ij}(x,y) = 0$.
Wigner transforming and taking up to the second-order terms in the derivative expansion, we obtain
\begin{eqnarray}
\Sigma_K (X;P) &\simeq& 
i\tilde{\Gamma}^{(p)} + i\left\{ f, Re\Sigma_R \right\} 
-\frac{1}{8} \left\{ \left\{ f, \Sigma_{12} - \Sigma_{21} \right\} \right\},
\label{eq:Sigma_K(X;P)}\\
i\tilde{\Gamma}^{(p)}(X;P) &\equiv& (1+f) \Sigma_{12} - f \Sigma_{21},
\label{eq:itildeGamma(X;P)}
\end{eqnarray}
where use has been made of $\Sigma_A(X;P) = \left[ \Sigma_R(X;P) \right]^\ast$.
$\tilde{\Gamma}^{(p)}$ is proportional to the net production rate of quasiparticles (see below after Eq.(\ref{eq:[[N,S]]})).
Eq.(\ref{eq:ulhatG}) with Eqs.(\ref{eq:hatDelta_RA}) and (\ref{eq:ulhatSigma(x,y)}) tells us that $\underline{\hat{G}}$ is of the form,
\begin{equation}
\underline{\hat{G}}(x,y) = 
\left( 
\begin{array}{cc}
G_R  &  G_K  \\
0  &  - G_A
\end{array}
\right)
\label{eq:ulhatG(x,y)}
\end{equation}
with
\begin{eqnarray}
G_{R(A)}(x,y) &=& 
\Delta_{R(A)} + \Delta_{R(A)} \cdot \Sigma_{R(A)} \cdot G_{R(A)},
\label{eq:G_R/A(x,y)}\\
G_K(x,y) &=& - G_R \cdot \Sigma_K \cdot G_A.
\label{eq:G_K(x,y)}
\end{eqnarray}
Wigner transforming and solving Eq.(\ref{eq:G_R/A(x,y)}), we obtain
\begin{eqnarray}
G_{R}(X;P) &=& 
\left[ G_A(X;P) \right]^\ast \simeq 
G_{R}^{(0)}(X;P) + G_{R}^{(2)}(X;P),
\label{eq:G_R/A(X;P)}\\
G_{R}^{(0)}(X;P) &=& 
\frac{1}{P^2 - \mathcal{M}^2(X;p_0) - \Sigma_{R}(X;P) + i\epsilon(p_0)0^+},
\label{eq:G_R/A^(0)(X;P)}\\
G_{R}^{(2)}(X;P) &=& 
\frac{\mathcal{G}_{R}^{\prime\prime}(X;P) \left( \Delta_{R}^{(0)}(X;P) \right)^{-1}}
{P^2 - \mathcal{M}^2(X;p_0) - \Sigma_{R}(X;P) + i\epsilon(p_0) 0^+},
\label{eq:G_R/A^(2)(X;P)}
\end{eqnarray}
where
\begin{eqnarray}
\mathcal{G}_{R}^{\prime\prime}(X;P) 
&=& 
{\Delta_R^{(0)}}^{-1} \Delta_R^{(2)} G_R^{(0)}
- \frac{1}{4} \left\{ \Delta_{R}^{(0)}, \left\{ \Sigma_{R}, G_{R}^{(0)} \right\} \right\}
\nonumber\\
& &- \frac{1}{8} \Delta_{R}^{(0)} 
\left\{ \left\{ \Sigma_{R}, G_{R}^{(0)} \right\} \right\}
-\frac{1}{8} \left\{ \left\{ \Delta_{R}^{(0)}, \Sigma_{R} G_{R}^{(0)} \right\} \right\}.
\label{eq:calG_R/A^''(X;P)}
\end{eqnarray}
We may write $G_R(X;P)$ in the form,
\begin{equation}
G_R(X;P)
=
\frac{1}{ 
P^2 - \mathcal{M}^2(X;p_0) - \Sigma_R(X;P) - \mathcal{F}(X;P)
},
\label{eq:G_Ragain}
\end{equation}
where
\begin{equation}
\mathcal{F}(X;P) 
= 
\left[
\Delta_R^{(0)}(X;P) G_R^{(0)}(X;P)
\right]^{-1} 
\mathcal{G}_R^{\prime\prime}(X;P).
\label{eq:calF}
\end{equation}
For $G_K(X;P)$, we obtain from Eq.(\ref{eq:G_K(x,y)}), after some manipulation,
\begin{eqnarray}
G_K(X;P) &\simeq& 
- \Sigma_K G_R^{(0)} G_A^{(0)}
\nonumber\\
& &- 
\frac{1}{2i} G_R^{(0)} G_A^{(0)}
\Bigg(  
\Sigma_K G_R^{(0)} G_A^{(0)}
\left[  
\left\{ \Sigma_R, \Sigma_A \right\} 
+ \left\{ P^2- \mathcal{M}^2, \Sigma_R - \Sigma_A \right\} 
\right] 
\nonumber\\
& &+
\partial_{P_\mu} \Sigma_K
\left[  
\left( \partial_{X^\mu}\Sigma_R + \partial_{X^\mu} \mathcal{M}^2 \right) 
G_R^{(0)}
- \left( \partial_{X^\mu}\Sigma_A + \partial_{X^\mu} \mathcal{M}^2 \right) 
G_A^{(0)}
\right]
\nonumber\\
& &-
\partial_{X^\mu} \Sigma_K 
\left[ 
\left( \partial_{P_\mu}\Sigma_R - 2P^\mu \right) G_R^{(0)} 
- \left( \partial_{P_\mu}\Sigma_A - 2P^\mu \right) G_A^{(0)}
\right]
\Bigg) .
\label{eq:G_K(X;P)}
\end{eqnarray}
In Eq.(\ref{eq:G_K(X;P)}), we have kept up to the first-order term of the derivative expansion.
The reason is as follows: 
In the physical-$N$ scheme, which will be constructed in Sec.\ref{sec:Boltzmann} below, $\Sigma_K$ is proportional to $\partial N^\pm / \partial X^\mu$, with $N^+ (N^-)$ the ``physical'' number density of the quasiparticle (anti-quasiparticle), and already is of the same order of magnitude as the first-order term of the derivative expansion.
As has been shown in~\cite{Niegawa} and as will be briefly discussed in Secs.\ref{sec:bareNscheme} and \ref{sec:Discussion} below, the bare-$N$ scheme is equivalent to the physical-$N$ scheme in the sense that they lead to the same result for the physical quantities.
Then, the above statement applies also to the bare-$N$ scheme.

We are now in a position to obtain a form for $\hat{G}(X;P)$ from Eq.(\ref{eq:haha}).
Straightforward but tedious manipulation yields
\begin{eqnarray}
\hat{G}(X;P) &\simeq& 
\hat{G}^{(0)}(X;P) + \hat{G}^{(1)}(X;P) + \hat{G}^{(2)}(X;P) 
+ G_K(X;P) \hat{A}_+,
\label{eq:hatG012K}\\
\hat{G}^{(0)}(X;P) &=& 
\left( 
\begin{array}{cc}
G_R^{(0)}  &  0  \\
G_R^{(0)} - G_A^{(0)}  &  - G_A^{(0)}
\end{array}
\right)
+ \left[ \left( G_R^{(0)} - G_A^{(0)} \right) f \right] \hat{A}_+,
\label{eq:hatG0}\\
\hat{G}^{(1)}(X;P) &=&
\frac{1}{2i}\left\{ G_R^{(0)} + G_A^{(0)} , f \right\} \hat{A}_+,
\label{eq:hatG1}\\
\hat{G}^{(2)}(X;P) &=&
\left( 
\begin{array}{cc}
G_R^{(2)}  &  0  \\
G_R^{(2)} - G_A^{(2)}  &  - G_A^{(2)}
\end{array}
\right)
\nonumber\\
& &+ \left[ \left( G_R^{(2)} - G_A^{(2)} \right) f \right] \hat{A}_+
- \frac{1}{8}\left\{ \left\{ G_R^{(0)} - G_A^{(0)} , f \right\} \right\} 
\hat{A}_+,
\label{eq:hatG2}
\end{eqnarray}
where
\begin{eqnarray}
\left\{ G_R^{(0)} + G_A^{(0)} , f \right\} &=& 
\left\{ f, P^2 - \mathcal{M}^2 - \Sigma_R \right\} {G_R^{(0)}}^2 
\nonumber \\
& &+ \left\{ f, P^2 - \mathcal{M}^2 - \Sigma_A \right\} {G_A^{(0)}}^2,
\label{eq:[G_R+G_A,f]}\\
\left\{ \left\{ G_R^{(0)} - G_A^{(0)} , f \right\} \right\} &=& 
\left\{ \left\{ -P^2 + \mathcal{M}^2 +\Sigma_R, f \right\} \right\} 
{G_R^{(0)}}^2
\nonumber\\
& &+
\Big[
-4 \left( \partial_{X^\mu}\Sigma_R + \partial_{X^\mu}\mathcal{M}^2 \right)
   \left( \partial_{P_\nu}\Sigma_R - 2P^\nu \right)
   \partial_{P_\mu}\partial_{X^\nu}f
\nonumber\\
& &
+2 \left( \partial_{X^\mu}\Sigma_R + \partial_{X^\mu}\mathcal{M}^2 \right)
   \left( \partial_{X^\nu}\Sigma_R + \partial_{X^\nu}\mathcal{M}^2 \right)
   \partial_{P_\mu}\partial_{P_\nu}f
\nonumber\\
& &
+2 \left( \partial_{P_\mu}\Sigma_R - 2P^\mu \right)
   \left( \partial_{P_\nu}\Sigma_R - 2P^\nu \right)
   \partial_{X^\mu}\partial_{X^\nu}f
\Big] {G_R^{(0)}}^3
\nonumber\\
& &- \left( R \rightarrow A \right).
\label{eq:[[G_R-G_A,f]]}
\end{eqnarray}
Inspection of Eqs.(\ref{eq:hatG012K})--(\ref{eq:[[G_R-G_A,f]]}) shows that all but $G_K\hat{A}_+$ do not bring about any pathology into perturbation theory.
We observe that $G_K$ in Eq.(\ref{eq:G_K(X;P)}) includes the product $G_R^{(0)}(X;P) G_A^{(0)}(X;P)$.
As seen from Eq.(\ref{eq:G_R/A^(0)(X;P)}), in the narrow width approximation, $Im\Sigma_R (= -Im\Sigma_A) \rightarrow -\epsilon(p_0)0^+$, $G_R^{(0)} G_A^{(0)}$ develops pinch singularity in a complex $p_0$-plane.
Then the contribution of $G_K(X;P)$ to some amplitude diverges on the energy shell, $ReG_R^{-1}(X;P) = 0$ (c.f. Eq.(\ref{eq:G_Ragain})).
In practice, $ImG_R^{-1} (\propto \lambda^2)$ is a small quantity, so that the contribution , although not divergent, is large.
This invalidates the perturbative scheme.

In the next two sections, we present two perturbative schemes, the bare-$N$ scheme and the physical-$N$ scheme.

\section{Bare-$N$ scheme}\label{sec:bareNscheme}
In this section, we briefly describe the bare-$N$ scheme~\cite{Niegawa}. \

\subsection{Preliminary}\label{ssec:NumberDensity}
In order to clarify the physical meaning of $f$, we compute the statistical average of current density,
\begin{equation}
j^\mu(x) \equiv 
\frac{i}{2} 
\left[ 
\phi^\dagger(x) 
\stackrel{\leftrightarrow}{\partial^\mu}
\phi(x)
- \phi(x)
\stackrel{\leftrightarrow}{\partial^\mu}
\phi^\dagger(x)
\right],
\label{eq:j^mu(x)}
\end{equation}
where $\stackrel{\leftrightarrow}{\partial^\mu} = \partial^\mu - \overleftarrow{\partial^\mu}$.
Taking the statistical average of Eq.(\ref{eq:j^mu(x)}), we obtain
\begin{equation}
\langle j^\mu(X) \rangle = 
\int \frac{d^4P}{(2\pi)^4} iP^\mu G_c(X;P),
\label{eq:<j^mu(X)>}
\end{equation}
where $G_c(X;P) \equiv G_{11} + G_{22} = G_{12} + G_{21}$.
Now we first consider the contribution from $G_c^{(0)}(X;P) = G_{11}^{(0)} + G_{22}^{(0)}$, the leading part of $G_c(X;P)$ in the derivative expansion.
When the interaction is switched off, $G_c^{(0)}$ reduces to $\Delta_c^{(0)} = \Delta_{11}^{(0)} + \Delta_{22}^{(0)}$ with $\mathcal{M}^2(X;p_0) = m^2$.
The contribution from $\Delta_c^{(0)}$ with $\mathcal{M}^2(X;p_0) \rightarrow m^2$ (see Eqs.(\ref{eq:hatDelta0}) with Eq.(\ref{eq:Delta_R/A^(0)})) is 
\begin{equation}
\langle j^0(X) \rangle - \langle 0 | j^0(X) | 0 \rangle = 
\int \frac{d^3P}{(2\pi)^3}
\left[ 
f(X;E_p,\mathbf{p}) + \left\{ 1 + f(X;-E_p,\mathbf{p}) \right\} 
\right],
\label{eq:<j^0(X)>}
\end{equation}
\begin{equation}
\langle \mathbf{j}(X) \rangle - \langle 0 | \mathbf{j}(X) | 0 \rangle = 
\int \frac{d^3P}{(2\pi)^3} \frac{\mathbf{p}}{E_p}
\left[ 
f(X;E_p,\mathbf{p}) - \left\{ 1 + f(X;-E_p,\mathbf{p}) \right\} 
\right],
\label{eq:<bfj(X)>}
\end{equation}
where $E_\mathbf{p} = \sqrt{\mathbf{p}^2 + m^2}$, $f(X;\pm E_\mathbf{p},\mathbf{p}) = f(X;P)|_{p_0=\pm E_\mathbf{p}}$.
These are to be compared with 
\begin{equation}
\langle j^\mu(X) \rangle - \langle 0 | j^\mu(X) | 0 \rangle = 
\int \frac{d^3P}{(2\pi)^3} v^\mu
\left[ 
N^+(X;E_\mathbf{p},\hat{\mathbf{p}}) - N^-(X;E_\mathbf{p},\hat{\mathbf{p}}) 
\right],
\label{eq:<j^mu(X)>-<0>}
\end{equation}
where $v^\mu = \left( 1, \mathbf{p}/E_\mathbf{p} \right)$ is the four-velocity and $N^+(X;E_\mathbf{p},\hat{\mathbf{p}})~~[N^-(X;E_\mathbf{p},\hat{\mathbf{p}})]$ is the number density of the quasiparticle [anti-quasiparticle] with momentum $\mathbf{p}$.
Comparing Eqs.(\ref{eq:<j^0(X)>}) and (\ref{eq:<bfj(X)>}) with Eq.(\ref{eq:<j^mu(X)>-<0>}), we have
\begin{equation}
f(X;p_0=E_\mathbf{p},\mathbf{p}) = N^+(X;E_\mathbf{p},\hat{\mathbf{p}}), ~~~~~
f(X;p_0=-E_\mathbf{p},\mathbf{p}) = - 1 - N^-(X;E_\mathbf{p},-\hat{\mathbf{p}})
.
\label{eq:ftoN}
\end{equation}
Thus, the physical number densities are obtained through computing Eq.(\ref{eq:<j^mu(X)>}).
Eq.(\ref{eq:<j^mu(X)>}) with $G_c^{(0)}$ for $G_c$ leads to a perturbative correction to Eqs.(\ref{eq:<j^0(X)>}) and (\ref{eq:<bfj(X)>}), due to quantum and medium effect.
This is also the case for $G_c^{(1)}$ and $G_c^{(2)}$ (c.f. Eqs.(\ref{eq:hatG1}) -- (\ref{eq:[[G_R-G_A,f]]})).
On the contrary, as has been discussed at the end of Sec.~\ref{sec:Resum}, the contribution from $G_K(X;P) \left( \in G_c(X;P) \right)$ is disastrously large.

\subsection{Bare-$N$ scheme}\label{ssec:Bare-N}
In this subsection, we write $f^\mathrm{(B)}$ for $f$ above.
The ``bare'' number densities $N^{\mathrm{(B)}+}(X;p_0,\hat{\mathbf{p}}) = \theta(p_0) f^\mathrm{(B)}(X;\mathbf{p})$ and $N^{\mathrm{(B)}-}(X;|p_0|,\hat{\mathbf{p}}) = -1 - \theta(-p_0)f^\mathrm{(B)}(X;p_0,-\mathbf{p})$ obey the ``free Boltzmann equation'', Eq.(\ref{eq:freeBoltz}) with $f^\mathrm{(B)}$ for $f$, which is to be solved under the given initial data $f^\mathrm{(B)}(X_i^0, \mathbf{X};P) = f(X_i^0, \mathbf{X};P)$ with $X_i^0$ the initial time.

The physical number densities, which are obtained from $\langle j^\mu(X) \rangle$, Eq.(\ref{eq:<j^mu(X)>}), are functionals of $f^\mathrm{(B)}$:
\begin{eqnarray}
f^{\mathrm{(ph)}}(X;P) 
&=& 
\theta(p_0) N^{\mathrm{(ph)}+}(X;p_0, \mathbf{p})
+ \theta(-p_0)
\left[
1 - N^{\mathrm{(ph)}-}(X;|p_0|,-\mathbf{p})
\right]
\nonumber \\
&=&
\mathcal{H}(X;P;[f^{\mathrm{(B)}}]).
\label{eq:H(f^B)}
\end{eqnarray}
$\mathcal{H}$ here contains large contributions mentioned above.
Solving this equation for $f^\mathrm{(B)}$, one obtains
\begin{equation}
f^{\mathrm{(B)}} = f^{\mathrm{(B)}}(X;P;[f^{\mathrm{(ph)}}]).
\label{eq:f^Boff^ph}
\end{equation}
Computation of some physical quantity yields the expression $F([f^{\mathrm{(B)}}])$, which includes large contribution.
Substituting the RHS of Eq.(\ref{eq:f^Boff^ph}) for $f^{\mathrm{(B)}}$ in $F$, one obtains the expression $F^\prime([f^{\mathrm{(ph)}}])$, which does not include large contribution.
The perturbation theory thus constructed is called the ``bare-$N$ scheme'' in~\cite{Niegawa}.

\section{Physical-$N$ scheme, and Gap and Boltzmann equations}\label{sec:Boltzmann}
In this section, we construct a ``physical-$N$ scheme'' in terms of the number densities that are as close as possible to the physical number density.
To this end, introducing a new function $f(x,y)$, we redefine $\hat{\Delta}(x,y)$ by Eq.(\ref{eq:hatDelta(x,y)}) with this new $f(x,y)$ in $\hat{B}_{L(R)}$.
Then, the ``free Boltzmann equation'' (\ref{eq:freeBoltz}) does not hold.
Specification of $f$ will be made in Sec.~\ref{ssec:Boltzmann}.

Now, in contrast to Eq.(\ref{eq:yama}), $\hat{\Delta}$ is not an inverse of $-\hat{\tau}_3 \left( \delta^4(x-y) \partial_y^2 + \mathcal{M}^2(x,y) \right)$, so that $\hat{\mathcal{L}}_0$ in Eq.(\ref{eq:hatLag_0}) is not the free hat-Lagrangian.
For the purpose of finding the free hat-Lagrangian, we first compute $\int d^4z \hat{\tau}_3 (\delta^4(x-z) \partial_z^2 + \mathcal{M}^2(x,z)) \hat{\Delta}(z,y)$ with $\hat{\Delta}(z,y)$ as in Eq.(\ref{eq:hatDelta(x,y)}):
\begin{eqnarray}
& &\int d^4z 
\hat{\tau}_3 \left( \delta^4(x-z) \partial_z^2 + \mathcal{M}^2(x,z) \right) 
\hat{\Delta}(z,y) 
\nonumber\\
& &~~~~
= 
-\delta^4(x-y)
\nonumber \\
& &~~~~~~~
+ 
\hat{\tau}_3 
\Biggl[
\Biggl\{  
\left( 
\partial^2 \hat{B}_L
- \hat{B}_L \stackrel{\leftarrow}{\partial^2} 
\right) 
+
\mathcal{M}^2 \cdot \hat{B}_L - \hat{B}_L \cdot \mathcal{M}^2 
\Biggr\} \cdot \hat{B}_L^{-1} \cdot \hat{\Delta}
\Biggr] (x,y)
,
\label{eq:TauKerDelta}
\end{eqnarray}
where we have used the short-hand notation (\ref{eq:dotSeki}).
Substituting Eq.(\ref{eq:hatB_L(x,y)}) for $\hat{B}_L$ in $\hat{\Delta}$ in  Eq.(\ref{eq:TauKerDelta}), we obtain
\begin{eqnarray}
\int d^4z \hat{\mathcal{D}}(x,z) \hat{\Delta}(z,y) &\simeq& \delta^4(x-y),
\label{eq:hatDDelta}\\
\hat{\mathcal{D}}(x,y) &=& 
- \hat{\tau}_3 \left( \delta^4(x-y) \partial_y^2 + \mathcal{M}^2(x,y) \right)
\nonumber\\
& &+
i \int \frac{d^4P}{(2\pi)^4} e^{-iP \cdot (x-y)}
\left\{ f, P^2 - \mathcal{M}^2(X;p_0) \right\} \hat{A}_-
.
\label{eq:hatD(x,y)}
\end{eqnarray}
Eq.(\ref{eq:hatDDelta}) tells us that the free hat-Lagrangian is 
\begin{equation}
\hat{\mathcal{L}}_0^\prime = 
\int d^4y \hat{\phi}^\dagger(x) \hat{\mathcal{D}}(x,y) \hat{\phi}(y).
\label{eq:NewFreeL}
\end{equation}
Since $\hat{\mathcal{L}}_0^\prime \neq \hat{\mathcal{L}}_0$, Eqs.(\ref{eq:NewFreeL}) and (\ref{eq:hatLag_0}), the counter hat-Lagrangian $\hat{\mathcal{L}}_0 - \hat{\mathcal{L}}_0^\prime = \hat{\mathcal{L}}_{c2}$ appears:
\begin{equation}
\hat{\mathcal{L}}_{c2}
=
- \hat{\phi}^\dagger(x) 
\int d^4y 
\int \frac{d^4P} {(2\pi)^4} e^{-iP \cdot (x-y)}i
\left\{
f, P^2 - \mathcal{M}^2(X;p_0)
\right\} \hat{A}_- \hat{\phi}(y).
\label{eq:Lc2}
\end{equation}
$\hat{\mathcal{L}}_{c2}$ yields the 2-point vertex function 
$i\hat{V}_{c2} \equiv \left\{ f, P^2 - \mathcal{M}^2(X;p_0) \right\} \hat{A}_-$.
Thus, the theory contains three types of vertices, $i\hat{V}$ (Eq.(\ref{eq:Vertex})), $i\hat{V}_{c1}$ (Eq.(\ref{eq:Vertex_c1})), and $i\hat{V}_{c2}$.
In particular, $i\hat{V}_{c1}$ and $i\hat{V}_{c2}$ contribute to the self-energy $\hat{\Sigma}$, 
\begin{equation}
\hat{\Sigma} = 
\hat{\Sigma}^{\mathrm{(loop)}} 
+ (m^2 - \mathcal{M}^2(X;p_0)) \hat{\tau}_3
+ i \left\{ f, P^2 - \mathcal{M}^2(X;p_0) \right\} \hat{A}_-.
\label{eq:loop-Vc2}
\end{equation}
Here $\hat{\Sigma}^{\mathrm{(loop)}}$ is the contribution from the loop diagrams.

\subsection{Gap equation}\label{ssec:Gap}
For the purpose of determining so far arbitrary mass function $\mathcal{M}(X;p_0)$, different prescriptions are available (see, e.g.,~\cite{Hatsuda}.)
Here we employ the pole prescription,
\begin{equation}
Re \left[ G_R^{-1}(X;P) \right] 
\biggr|_{p_0=\pm M_\pm(X), \mathbf{p}=0} = 0,
\label{eq:Mrenorm}
\end{equation}
with $G_R(X;P)$ as in Eq.(\ref{eq:G_Ragain}).
Incorporating the contribution from the counter Lagrangian $\hat{\mathcal{L}}_{c1}$ (see Eq.(\ref{eq:loop-Vc2})), Eq.(\ref{eq:Mrenorm}) becomes
\begin{equation}
\left[ 
M_\pm^2(X) - m^2 
-Re \Sigma_R^{\mathrm{(loop)}}(X;P) 
- 
Re\mathcal{F}(X;P)
\right]
\biggr|_{p_0=\pm M_\pm(X), \mathbf{p}=0} 
= 0,
\label{eq:Tadpole=0}
\end{equation}
where $\mathcal{F}(X;P)$ is as in Eq.(\ref{eq:calF}).
Eq.(\ref{eq:Tadpole=0}) is the gap equation, from which $M_\pm^2(X)$ is determined self consistently.
It should be noted that, in Eq.(\ref{eq:Mrenorm}), instead of choosing $\mathbf{p}=0$, one can choose $\mathbf{p}=\mathbf{p}_c (\neq 0)$.

\subsection{Boltzmann equation}\label{ssec:Boltzmann}
We now define $f$, such that the number densities (c.f. Eq.(\ref{eq:ftoN}))
\begin{eqnarray}
N^+(X;p_0=\omega_+,\hat{\mathbf{p}}) 
&=& 
f(X;p_0=\omega_+,\hat{\mathbf{p}}),
\label{eq:N^+(X;omega+,hatbfp)}\\
N^-(X;-p_0=\omega_-,-\hat{\mathbf{p}}) 
&=& 
-1 -f(X;p_0=-\omega_-,\hat{\mathbf{p}})
\label{eq:N^-(X;omega-,hatbfp)},
\end{eqnarray}
are as close as possible to the physical-number densities.
Here $\omega_\pm = \omega_\pm(X;\pm\mathbf{p})$ is the energy of the quasiparticle (anti-quasiparticle) mode with momentum $\pm\mathbf{p}$:
\begin{eqnarray}
& &Re \left[ G_R^{-1}(X;P) \right]\biggr|_{p_0=\pm\omega_\pm(X;\pm\mathbf{p})}
\nonumber\\
& &=
\left[ 
P^2 - m^2
-Re \Sigma_R^{(\mathrm{loop})}(X;P) 
- Re \mathcal{F}(X;P)
\right]
\biggr|_{p_0=\pm\omega_\pm(X;\pm\mathbf{p})} = 0.
\label{eq:onshell}
\end{eqnarray}
As has been stressed in Sec.~\ref{ssec:NumberDensity}, $G_K$ in Eq.(\ref{eq:G_K(X;P)}), which includes $\Sigma_K$ (see Eq.(\ref{eq:G_K(x,y)})), yields a large contribution to the physical number density.
Then, as the determining equation for $f$, we adopt
\begin{equation}
\Sigma_K (X;P) = 0.
\label{eq:Sigma_K=0}
\end{equation}
It should be emphasized here that this condition is by no means unique~\cite{Niegawa}. \ 
As a matter of fact, no large contribution emerges, as long as we impose the condition (\ref{eq:Sigma_K=0}) on the energy shells and their vicinities.

Recalling Eq.(\ref{eq:loop-Vc2}) with Eq.(\ref{eq:Sigma_K(x,y)}), the condition (\ref{eq:Sigma_K=0}) turns out to
\begin{equation}
\Sigma_K^{\mathrm{(loop)}}(X;P) 
- i \left\{ f, P^2 - m^2 \right\}
= 0.
\label{eq:Sigma_K'=0}
\end{equation}
Performing the derivative expansion and taking up to second-order terms, Eq.(\ref{eq:Sigma_K'=0}) with Eq.(\ref{eq:Sigma_K(X;P)}) becomes
\begin{eqnarray}
2 P^\mu \partial_{X^\mu} f(X;P)
&-& 
\left\{ f(X;P), \mathcal{M}^2(X;p_0) + Re\Sigma_R(X;P) \right\} 
\nonumber \\
&=&
\tilde{\Gamma}^{(p)}_{\mathrm{(loop)}}(X;P)
\nonumber \\
& &+ \frac{i}{8} 
\left\{ \left\{ 
f(X;P), 
\Sigma_{12}^{\mathrm{(loop)}}(X;P) - \Sigma_{21}^{(\mathrm{loop})}(X;P) 
\right\} \right\}.
\label{eq:Boltzmann}
\end{eqnarray}

We are now in a position to inspect the physical implication of Eq.(\ref{eq:Boltzmann}).
Straightforward manipulation using Eqs.(\ref{eq:N^+(X;omega+,hatbfp)}) and (\ref{eq:N^-(X;omega-,hatbfp)}) (see Appendix \ref{sec:Derivation}) shows that Eq.(\ref{eq:Boltzmann}) becomes, on the energy-shells, $p_0=\pm\omega_\pm$,
\begin{eqnarray}
\frac{\partial N^\pm}{\partial X^0}
&+&
\mathbf{v}_\pm \cdot \nabla_\mathbf{X} N^\pm
\pm
\frac{\partial \omega_\pm}{\partial X^\mu}
\frac{\partial N^\pm}{\partial P_\mu}
\pm
\frac{Z_\pm}{2\omega_\pm}
\left[
\frac{\partial N^\pm}{\partial X^\mu}
\frac{\partial Re \mathcal{F}}{\partial P_\mu}
-
\frac{\partial N^\pm}{\partial P_\mu}
\frac{\partial Re \mathcal{F}}{\partial X^\mu}
\right]
\nonumber\\
&=&
Z_\pm
\left[  
\Gamma_\pm^{(p)} 
+ 
\frac{i}{16\omega_\pm} 
\left\{ \left\{ 
\pm N^\pm(X; \pm \omega_\pm, \pm \mathbf{p}), \Sigma_- (X;P)
\right\} \right\}
\right]_{p_0 = \pm \omega_\pm},
\label{eq:kinetic}
\end{eqnarray}
where $\mathbf{v}_\pm (= \pm \partial \omega_\pm / \partial \mathbf{p})$ is the group velocity of the $\pm$ mode with momentum $\pm \mathbf{p}$, $Z_\pm$ is the wave-function renormalization factor (cf. Eq.(\ref{eq:Z^-1})), and 
$\mathcal{F}$ is as in Eq.(\ref{eq:calF}).
In Eq.(\ref{eq:kinetic}),
\begin{eqnarray}
\Gamma_\pm^{(p)} 
&=& 
\frac{-i}{2\omega_\pm}
\left[  
\left( 1 + N^\pm \right) \Sigma_{12(21)}^{(\mathrm{loop})} 
- N^\pm \Sigma_{21(12)}^{(\mathrm{loop})}
\right]_{p_0 = \pm \omega_\pm},
\label{eq:Prate}\\
\Sigma_-(X;P) 
&\equiv& 
\Sigma_{12}^{(\mathrm{loop})}(X;P) - \Sigma_{21}^{(\mathrm{loop})}(X;P),
\label{eq:Sigma-}\\
\left\{ \left\{ 
\pm N^\pm, \Sigma_- 
\right\} \right\}
&\equiv&
\pm 
\frac{d^2N^\pm}{dX^\mu dX^\nu}
\frac{\partial^2 \Sigma_-}{\partial P_\mu \partial P_\nu}
\mp 2
\frac{d^2N^\pm}{dX^\mu dP_\nu}
\frac{\partial^2 \Sigma_-}{\partial P_\mu \partial X^\nu}
\pm 
\frac{d^2N^\pm}{dP_\mu dP_\nu}
\frac{\partial^2 \Sigma_-}{\partial X^\mu \partial X^\nu}
\nonumber\\
& &-n^\pm_0
\left[  
\frac{\partial^2 \omega_\pm}{\partial X^\mu \partial X^\nu}
\frac{\partial^2 \Sigma_-}{\partial P_\mu \partial P_\nu}
+
\frac{\partial^2 \omega_\pm}{\partial P_i \partial P_j}
\frac{\partial^2 \Sigma_-}{\partial X^i \partial X^j}
-2
\frac{\partial^2 \omega_\pm}{\partial X^\mu \partial P_j}
\frac{\partial^2 \Sigma_-}{\partial P_\mu \partial X^j}
\right]
\nonumber\\
& & \pm \frac{\partial n^\pm_0}{\partial p^0}
\left[  
\frac{\partial \omega_\pm}{\partial X^\mu}
\frac{\partial \omega_\pm}{\partial X^\nu}
\frac{\partial^2 \Sigma_-}{\partial P_\mu \partial P_\nu}
+
v_\pm^i v_\pm^j
\frac{\partial^2 \Sigma_-}{\partial X^i \partial X^j}
+2
\frac{\partial \omega_\pm}{\partial X^\mu}
v_\pm^j
\frac{\partial^2 \Sigma_-}{\partial P_\mu \partial X^j}
\right]
\nonumber\\
& &\mp 2v_\pm^i
\left[  
\frac{\partial^2 \Sigma_-}{\partial X^i \partial P_\nu}
\frac{dn^\pm_0}{dX^\nu}
-
\frac{\partial^2 \Sigma_-}{\partial X^i \partial X^\nu}
\frac{dn^\pm_0}{dP_\nu}
\right]
\nonumber\\
& &-2\frac{\partial \omega_\pm}{\partial X^\mu}
\left[  
\frac{\partial^2 \Sigma_-}{\partial P_\mu \partial P_\nu}
\frac{dn^\pm_0}{dX^\nu}
-
\frac{\partial^2 \Sigma_-}{\partial P_\mu \partial X^\nu}
\frac{dn^\pm_0}{dP_\nu}
\right],
\label{eq:[[N,S]]}
\end{eqnarray}
where $i$ and $j$ run over $1, 2, 3$ and $n_0^\pm \equiv \partial N^\pm / \partial p^0$.
$\Gamma_\pm^{(p)}$ that comes from $\tilde{\Gamma}_{(\mathrm{loop})}^{(p)}$ (Eq.(\ref{eq:itildeGamma(X;P)})) is the net production rate of the quasiparticle (anti-quasiparticle) with momentum $\mathbf{p} (-\mathbf{p})$.
In fact, $\Gamma_\pm^{(p)}$ is the difference between the production rate and the decay rate.

Several observations are in order.

1) The last term on the LHS and the last term on the RHS of Eq.(\ref{eq:kinetic}) are new, which comes from the second-order parts of the derivative expansion.

2) The third term on the LHS of Eq.(\ref{eq:kinetic}) comes from the force $\pm\partial \omega_\pm / \partial X^\mu$ due to the variation of the energy $\omega_\pm$ with $X^\mu$.

3) If the terms mentioned in 1) and 2) above are unimportant, and the renormalization of the wave function is neglected $(Z_\pm \simeq 1)$, Eq.(\ref{eq:kinetic}) becomes the standard transport equation,
\begin{equation}
\frac{\partial N^\pm}{\partial X^0} 
+ \mathbf{v}_\pm \cdot \nabla_\mathbf{X} N^\pm
=
\Gamma_\pm^{(p)}.
\label{eq:TransStd}
\end{equation}
Thus Eq.(\ref{eq:kinetic}) is a generalized transport or Boltzmann equation.

4) The last term on the LHS of Eq.(\ref{eq:kinetic}) is regarded as corrections to the first three terms:
\begin{eqnarray}
\mbox{LHS of Eq.(\ref{eq:kinetic})}
&=&
\left[
1 \pm \frac{Z_\pm}{2\omega_\pm} \frac{\partial Re \mathcal{F}}{\partial p_0}
\right]
\frac{\partial N^\pm}{\partial X^0}
+
\left[
\mathbf{v}_\pm \mp 
\frac{Z_\pm}{2\omega_\pm} \frac{\partial Re \mathcal{F}}{\partial \mathbf{p}}
\right]
\cdot \nabla_\mathbf{X} N^\pm
\nonumber\\
& &\pm
\left[
\frac{\partial \omega_\pm}{\partial X^\mu}
- \frac{Z_\pm}{2\omega_\pm} \frac{\partial Re \mathcal{F}}{\partial X^\mu}
\right]
\frac{\partial N^\pm}{\partial P_\mu}.
\label{eq:Left}
\end{eqnarray}
On the other hand the last term on the RHS of Eq.(\ref{eq:kinetic}) is regarded as a correction to the production rate, which comes from the fact that the spacetime point, at which a quasi-(anti)particle production (decay) occurs, does not coincide with the spacetime point of $N^\pm(X;\hat{\omega}_\pm,\pm \hat{\mathbf{p}})$.
(cf. Eq.(\ref{eq:Sigma_K(x,y)})).

5) On the light of the above observations, one can regard Eq.(\ref{eq:Boltzmann}) as an off the energy-shell generalization of the generalized Boltzmann equation.

Finally, it is worth mentioning that the physical number densities computed in the physical-$N$ scheme contain no large terms:
\begin{equation}
N^{\mathrm{(ph)}\pm}(X;P) = N^\pm(X;P) + \Delta N^\pm (X;P) +\dots ,
\label{eq:Aha}
\end{equation}
with $\Delta N^\pm$ the perturbative correction.

\subsection{Procedure of solving the gap and Boltzmann equation}\label{ssec:Perturbation}
The mass function $\mathcal{M}^2(X;p_0)$ and the number-density function $f(X;P)$ are determined by simultaneously solving the gap (\ref{eq:Tadpole=0}) and the Boltzmann (\ref{eq:Boltzmann}) equations.

As mentioned in Sec.~\ref{sec:Intro}, we are interested in quasiuniform systems near equilibrium and nonequilibrium quasistationary system.
Eq.(\ref{eq:TransStd}) tells us that the change of $N^\pm$ along the $(1, \mathbf{v}_\pm)$ is of $O(\lambda^2)$.
Then, it is natural to assume that $\partial_{X^\mu}f(X;P)$ is also of $O(\lambda^2)$.

One can solve this set of equations iteratively in perturbation theory.
Up to the third order of perturbation theory, the gap and the Boltzmann equations are dealt with to the gradient approximation, i.e., the last terms on the LHS of Eq.(\ref{eq:Tadpole=0}), and on the RHS of Eq.(\ref{eq:Boltzmann}) are dropped.

1) Zeroth order of perturbation theory.\\
The gap and Boltzmann equations become,
\begin{eqnarray}
M_{0\pm}^2(X) = m^2,
\label{eq:O0gap}\\
P^\mu \frac{\partial f_0(X;P)}{\partial X^\mu} = 0,
\label{eq:O0Boltz}
\end{eqnarray}
where $M_{0\pm}^2(X)$ and $f_0(X;P)$ are the $O(\lambda^0)=O(1)$, or ``bare'' parts of $M_\pm^2(X)$ and $f(X;P)$, respectively.
Solve this Boltzmann equation under the given initial data.

2) First order of perturbation theory.\\
To $O(\lambda)$, a tadpole diagram contributes to $\hat{\Sigma}^{(\mathrm{loop})} \equiv \left( \hat{\Sigma}^{(\mathrm{loop})(1)} \right)$.
It can readily be shown that $\Sigma_R^{(\mathrm{loop})(1)} \left( = \Sigma_A^{(\mathrm{loop})(1)} \right)$ is real and $P$-independent, and $\Sigma_{12}^{(\mathrm{loop})(1)} = \Sigma_{21}^{(\mathrm{loop})(1)} = 0$.
Then, the gap and the Boltzmann equations become
\begin{equation}
\left[ 
\Sigma_R^{(\mathrm{loop}) (1)}(X) - M_{1\pm}^2(X) 
\right] \biggr|_{p_0=\pm M_\pm(X),\mathbf{p}=0} 
= 0,
\label{eq:O1gap}
\end{equation}
and
\begin{equation}
2P^\mu \frac{\partial f_1(X;P)}{\partial X^\mu}
+ \frac{\partial \Sigma_R^{(1)}(X)}{\partial X^\mu}
\frac{\partial \left( f_0(X;P) + f_1(X;P) \right)}{\partial P_\mu}
= 0,
\label{eq:O1Boltz}
\end{equation}
respectively.
Here $M^2_{1\pm}(X)$ and $f_1(X;P)$ are the $O(\lambda)$ contributions to $M^2_{\pm}(X)$ and $f(X;P)$, respectively.
Incidentally, in computing $\Sigma_R^{(\mathrm{loop})(1)}(X)$, we may use the leading part $\hat{\Delta}^{(0)}(X;P)$ of the propagator $\hat{\Delta}(X;P)$ (see, Eq.(\ref{eq:hatDelta0})) with $\mathcal{M}^2 \rightarrow \mathcal{M}_{0}^2 + \mathcal{M}_{1}^2 \left( = m^2 + \mathcal{M}_1^2 \right)$ and $f \rightarrow f_0 + f_1$.
Eq.(\ref{eq:O1gap}) is immediately solved to give
\begin{equation}
M^2_{1+}(X) = M^2_{1-}(X) 
= \Sigma_R^{(\mathrm{loop})(1)}(X).
\label{eq:M+=M-}
\end{equation}
Eq.(\ref{eq:O1Boltz}) is to be solved under the initial data $f_1(X_{0in},\mathbf{X};P)=0$ with $X_{0in}$ the initial time.

3) Higher order calculation.\\
At the $O(\lambda^2)$ stage, $\tilde{\Gamma}_{(\mathrm{loop})}^{(p)}(X;P)$ in Eq.(\ref{eq:Boltzmann}) appears first.
At the $O(\lambda^3)$ stage, a new tadpole diagram appears, which is dealt with in a same manner as in the $O(\lambda)$ stage.
At the $O(\lambda^4)$ stage, the gap and the Boltzmann equations deduced in the present paper, by taking up to the second-order terms of derivative expansion, enters the stage.

\section{Discussion}\label{sec:Discussion}
Here we would like to mention a similarity between the two schemes presented here, the bare-$N$ scheme and the physical-$N$ scheme, and those in the ultra-violet renormalization schemes in quantum field theory.
For simplicity, we focus on the mass renormalization.

\subsection{Structure of UV-renormalization}\label{ssec:UVrenormalization}
\textit{``Bare'' UV-renormalization scheme}: \\
The free Lagrangian density reads $\mathcal{L}_0 = - \phi^\dagger(x) \left( \partial^2 + m_B^2 \right) \phi(x)$ with $m_B$ the bare mass.
Computation of the physical mass $m_\mathrm{ph}$ yields $m_\mathrm{ph} = m_\mathrm{ph}(m_B)$, which includes diverging terms.
Solving this equation for $m_B$, we obtain $m_B = m_B(m_\mathrm{ph})$.
Some physical quantity $F$, which we compute perturbatively, yields the expression $F = F(m_B)$, which contains, in general, UV-divergences.
Substituting the equation $m_B = m_B(m_\mathrm{ph})$ for $m_B$ in $F(m_B)$, one gets $F_R(m_\mathrm{ph}) \equiv F(m_B(m_\mathrm{ph}))$, which is free from UV-divergence. 

\textit{``Physical'' UV-renormalization scheme}: \\
We introduce new free Lagrangian $\mathcal{L}_0^\prime = - \phi^\dagger(x) \left( \partial^2 + m^2 \right) \phi(x)$ with $m$ the renormalized mass.
Then, the counter Lagrangian should be introduced, $\mathcal{L}_c = \mathcal{L}_m - \mathcal{L}_m^\prime = \phi^\dagger(x) \left[ m^2 - m_B^2 \right] \phi(x)$.
$m^2 - m_B^2$ is determined so that the perturbatively computed physical mass $m_\mathrm{ph}$ is free from the UV-divergences.
Thus, no diverging term is involved in the relation $m_\mathrm{ph} = m_\mathrm{ph}(m)$.

\subsection{Structure of the two schemes presented above}
\textit{Bare-$N$ scheme}: \\ 
The counter Lagrangian $\hat{\mathcal{L}}_{c2}$ is absent and $f$ obeys the ``free Boltzmann equation.''
The physical number densities, which are the functionals of $f^\mathrm{(B)}$, include large contributions.
Perturbative computation of some quantity yields the expression, which is written in terms of $f^\mathrm{(B)}$ and includes large contributions.
Rewriting it in terms of the physical quantity $f^\mathrm{(ph)}$, one obtains the large contribution free form.

\textit{Physical-$N$ scheme}: \\ 
We introduce a counter Lagrangian $\hat{\mathcal{L}}_{c2}$ (\ref{eq:Lc2}), which is determined so that the perturbatively computed physical number densities do not involve large contributions (cf. Eq.(\ref{eq:Aha})).

\subsection{Correspondence}\label{ssec:correspond}
Above observation discloses the correspondence between the two schemes presented here and those in the UV-renormalization scheme:

\textit{Bare scheme}: 
\begin{eqnarray}
\mathcal{L}_0 (x) &\leftrightarrow& \mathcal{L}_0 + \mathcal{L}_{c1}
~~~ \mbox{(Eq.(\ref{eq:LagC1}))}, \nonumber \\
m_\mathrm{B} &\leftrightarrow& f^\mathrm{(B)}, \nonumber \\
m_\mathrm{ph} &\leftrightarrow& f^\mathrm{(ph)}. \nonumber
\end{eqnarray}

\textit{Physical scheme}:
\begin{eqnarray}
\mathcal{L}_0^\prime (x) &\leftrightarrow& \hat{\mathcal{L}}_0^\prime
~~~ \mbox{(Eq.(\ref{eq:NewFreeL}))}, \nonumber \\
\mathcal{L}_c (x) &\leftrightarrow& \hat{\mathcal{L}}_{c2}
~~~ \mbox{(Eq.(\ref{eq:Lc2}))}, \nonumber \\
m &\leftrightarrow& f, \nonumber \\
m_\mathrm{ph} = m_\mathrm{ph} 
&\leftrightarrow& 
f^\mathrm{(ph)}=f^\mathrm{(ph)}[f] ~~~(\mbox{Eq.}(\ref{eq:Aha})), \nonumber \\
\mbox{Absence of }\nonumber \\
\mbox{divergence} &\leftrightarrow& \mbox{large contribution}. \nonumber
\end{eqnarray}

\section{Summary}\label{sec:Summary}
In this paper, for dealing with the out of equilibrium complex-scalar systems, we have constructed two kind of perturbation theories, the bare-$N$ scheme and the physical-$N$ scheme, beyond the gradient approximation in the derivative expansion.
The physical-$N$ scheme has been constructed by imposing the condition $\Sigma_K(x,y)=0$, Eq.(\ref{eq:Sigma_K'=0}).
This condition, which turns out to be a generalized Boltzmann equation on the energy-shell, serves as the determining equation for the number density function $f(X;P)$.
At the same time, the effective mass $M^2_\pm(X)$ is determined through the gap equation (\ref{eq:Tadpole=0}).

\section*{Acknowledgements}
I would like to thank Prof. A. Ni\'{e}gawa for useful comments, discussions and careful reading of the manuscript.
We thank the useful discussion at the Workshop on Thermal Quantum Field Theories and their Application, held at the Yukawa Institute for Theoretical Physics, Kyoto, Japan, 8 -- 10 August 2002.

\appendix
\section{Derivation of Eq.(\ref{eq:kinetic})}\label{sec:Derivation}
Here we derive Eq.(\ref{eq:kinetic}).
We define ``the on-shell energy'' $p_0 = \pm \omega_\pm(X;\pm\mathbf{p}) (\equiv \pm\omega_\pm)$ through Eq.(\ref{eq:onshell}), from which we obtain
\begin{eqnarray}
\frac{
\partial Re \left(\Sigma_R^{(\mathrm{loop})}(X;P) + \mathcal{F}(X;P) \right)
}{\partial \mathbf{p}}
&=& 
-2 \mathbf{p} \pm 2 \omega_\pm \mathbf{v}_{\pm} Z_\pm^{-1},
\label{eq:PDSigma}\\
\frac{
\partial Re \left(\Sigma_R^{(\mathrm{loop})}(X;P) + \mathcal{F}(X;P) \right)
}{\partial X^\mu}
&=&
2 \omega_\pm 
\frac{\partial \omega_\pm}{\partial X^\mu}
Z_\pm^{-1},
\label{eq:XDSigma}
\end{eqnarray}
where $\mathcal{F}(X;P)$ is as in Eq.(\ref{eq:calF}), $\mathbf{v}_\pm \equiv \pm \partial\omega_\pm / \partial \mathbf{p}$ is the group velocity of the $\pm$ mode with momentum $\pm \mathbf{p}$, and 
\begin{equation}
Z_\pm^{-1} \equiv 1 \mp \frac{1}{2\omega_\pm}
\frac{\partial Re (\Sigma_R^{(\mathrm{loop})} + \mathcal{F})}{\partial p^0}
\biggr|_{p_0 = \pm \omega_\pm}
\label{eq:Z^-1}
\end{equation}
is the wave-function renormalization factor.

The determining equation for $f$, Eq.(\ref{eq:Boltzmann}), is written in the form,
\begin{equation}
2 P^\mu \frac{\partial f}{\partial X^\mu}
- \frac{\partial f}{\partial X^\mu} 
  \frac{\partial Re \Sigma_R^{(\mathrm{loop})}}{\partial P_\mu}
+ \frac{\partial f}{\partial P_\mu}
  \frac{\partial Re \Sigma_R^{(\mathrm{loop})}}{\partial X^\mu}
=
\tilde{\Gamma}_{(\mathrm{loop})}^{(p)}
+ \frac{i}{8}
\left\{ \left\{ 
f, \Sigma_{12}^{(\mathrm{loop})} - \Sigma_{21}^{(\mathrm{loop})}
 \right\} \right\}.
\label{eq:detf}
\end{equation}
Putting Eqs.(\ref{eq:PDSigma}) and (\ref{eq:XDSigma}) into Eq.(\ref{eq:detf}), we obtain
\begin{eqnarray}
\left( \mbox{LHS of Eq.(\ref{eq:detf})} \right) 
&=&
2 p^0 \frac{\partial f}{\partial X^0}
+ 2 \mathbf{p} \cdot \nabla_\mathbf{X} f
- \frac{\partial f}{\partial X^0} 
\frac{\partial Re \Sigma_R^{(\mathrm{loop})}}{\partial P_0}
\nonumber\\
& &+ \frac{\partial f}{\partial \mathbf{X}}
\frac{\partial Re \Sigma_R^{(\mathrm{loop})}}{\partial \mathbf{P}}
+ \frac{\partial f}{\partial P_\mu}
\frac{\partial Re \Sigma_R^{(\mathrm{loop})}}{\partial X^\mu}
\nonumber\\
&\rightarrow& 
\pm 2 \omega_\pm Z_\pm^{-1} \frac{\partial f}{\partial X^0}
\pm 2 \omega_\pm Z_\pm^{-1} \mathbf{v}_{\pm} \cdot \nabla_\mathbf{X}f
+ 2 \omega_\pm Z_\pm^{-1} 
\frac{\partial f}{\partial P_\mu} \frac{\partial \omega_\pm}{\partial X^\mu}
\nonumber\\
& &+ \frac{\partial f}{\partial X^\mu}
\frac{\partial Re \mathcal{F}}{\partial P_\mu}
- \frac{\partial f}{\partial P_\mu}
\frac{\partial Re \mathcal{F}}{\partial X^\mu}.
\label{eq:detfL}
\end{eqnarray}
Multiplying $Z_\pm/2\omega_\pm$ to both sides of Eq.(\ref{eq:detf}) and using Eqs.(\ref{eq:N^+(X;omega+,hatbfp)}) and (\ref{eq:N^-(X;omega-,hatbfp)}), we obtain Eq.(\ref{eq:kinetic}).


\end{document}